\documentclass{appolbnh}
\usepackage{graphicx}
\usepackage{color}
\usepackage{lineno}
\begin{document}

\title{History of $N(1680)$ %
}

\author{Igor Strakovsky
\thanks{igor@gwu.edu}
\address{Institute for Nuclear Studies, Department of Physics, \\
         The George Washington University, Washington, DC 20052, USA}
}

\maketitle

\begin{abstract}
This paper describes my personal appreciation for some of the great research achievements of Mitya Diakonov, Vitya Petrov, and Maxim Polyakov and how my own research career has followed the paths they opened. Among the topics where they have been the most influential have been the pursuit and study of the exotic pentaquark. The search for exotics may require a complementary approach, such as experimental and theoretical activity. Here, I would like to focus on a story of $N(1680)$, a non-strange unitary partner of $\Theta^+$, in which I was involved.
\end{abstract}


\section*{Dedication}
This paper is dedicated to the memory of Mitya, Vitya, and Maxim. Perhaps now that they are no longer with us, we can better appreciate their scientific achievements.

\section{Prehistory}
I am an experimentalist and served in the Leningrad Nuclear Physics Institute (LNPI or PNPI now) High Energy Physics Division (HEPD) for 29 years. In particular, the reaction $\pi p\to p \pi\pi$ was an important part of the HEPD research. From my friends Senya Sherman (experimentalist) and Tolya Bolokhov (theorist), I had the opportunity to learn about an extraordinary student, Maxim, who was heavily involved in this activity, and he had several joint papers with seniors back to the end of 1980s. 

Somewhere in 1997 or so (the time of the famous $\overline{\textbf{10}}$ paper~\cite{Diakonov:1997mm}), Dick Arndt forwarded me a letter from Maxim asking what the Virginia Tech Scattering Analysis Interactive Dial-in (SAID) Kaon-nucleon Partial-Wave Analysis (PWA)~\cite{Hyslop:1992cs} gives for the exotic $S = +1$ baryon. Dick was very surprised, since the analysis gave nothing which may fit such a narrow exotic state. All four resonances in standard PWA are too heavy (mass $> 1790~\mathrm{MeV}$) and too broad (width $> 95~\mathrm{MeV}$). So they do not fit the predicted $\Theta^+$ in~\cite{Diakonov:1997mm}. This fact was a motivation for us to solve a standard PWA problem, and we do a modified PWA to look for narrow resonances~\cite{Arndt:2003xz}.

It was a good surprise for me to see that the famous $\overline{\textbf{10}}$ paper~\cite{Diakonov:1997mm} used our SAID sigma-term result~\cite{Pavan:2001wz}
for the mass difference in anti-decuplet using the Gell-Mann-Okubo phenomenological mass formula. 

The experimental results published in the top peer-review journals had problems for experimentalists and the main reason was English. I was a junior researcher at PNPI and was willing to write papers on $\pi^\pm p$ elastic scattering and proton-induced pion production to report the results of our group measurements at Gatchina.  I translated the Russian versions into English and asked Mitya to polish them.  Sure, Mitya ignored my silly staff, and one eve per paper was enough for him to make drafts which Nucl. Phis. A accepted in 1981 and then J. Phys. G accepted in 1988.  It will be good to acknowledge Mitya for his help in translation, but our experimental physics was far away from his mainstream, and for that reason Mitya's name is still hidden in both papers. 

In 2005 or so, Mitya gave a talk at a JLab seminar about pentaquark, and Dick Arndt, who was sitting next to me, asked where this guy got such English. My answer was simple - his mother (Nina Yakovlevna) was a professor of English literature in St.~Petersburg State University and translated a poem by John Keats from English into Russian. Dick was satisfied with my answer. 

I met Vitya for the first time at one of the LNPI Winter Schools at the end of the 1970s, but there is a background. In 1968, I measured the efficiency of neutron detection with large scintillation counters, using a beam of neutrons from the Neutron Generator (that was my Master project). At the same time, a physicist and an engineer studied the reaction of transistors to neutrons using the same source of fast neutrons. The physicist's name was Tamara Markovna. One day, a strange man appeared looking for his wife. Somebody told me who he was, Yurii Viktorovich Petrov, a well-known theoretician. That’s how I found Tamara Markovna's last name. 
They were Vitya's parents and Vitya was probably 13~years old at the time.

Obviously, the Instanton vacuum model developed by Mitya and Vitya is a very powerful
model~\cite{Diakonov:1985eg}. Before the $\overline{\textbf{10}}$ tasks came up, our friends focused on the nucleon-nucleon interaction and, in particular, on the dibaryon (see, for instance, Ref.~\cite{Diakonov:1988ju}).  In 1996 or 1997, I asked my postdoc, Serezha Pavlenko, to work with Mitya and Vitya because that was
a time when many groups were looking for dibaryon. He was a brilliant young experimentalist and phenomenologist. In a short time, Serezha did many calculations and Mitya and Vitya were impressed with the Serezha results. Unfortunately, he passed away shortly after the stroke. In 1998, Serezha was 24 years old.

\section{Introduction}
QCD gives rise to the hadron spectrum~\cite{Gell-Mann:1964ewy} and many $q\bar{q}$ and $qqq$ have been observed~\cite{ParticleDataGroup:2024cfk}. However, $q\bar{q}q\bar{q}$ and $qqqq\bar{q}$ and other many-quark states are not forbidden but have not yet been observed. Recently, the LHCb Collaboration claimed evidence for four hidden-charm $qqqc\bar{c}$ states near open-charm decay thresholds for $\Sigma_c^+\bar{D}^0$ and $\Sigma_c^+\bar{D}^{\ast 0}$ in $\Lambda_b^0 \to P^+_{c\bar{c}}K^- \to J/\psi p K^-$ decay~\cite{LHCb:2019kea}. However, although there is no doubt about the LHCb observations, it is not clear whether they are compact multiquark or hadonic molecular states~\cite{Eides:2017xnt}. 

\begin{figure}[htb!]
\centering
{
    \includegraphics[width=0.7\textwidth,keepaspectratio]{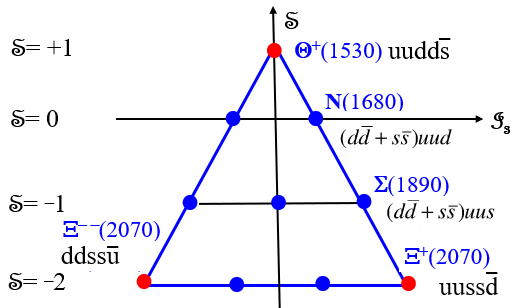} 
}

\centerline{\parbox{1\textwidth}{
\caption[] {\protect\small
The suggested anti-decuplet of baryons~\cite{Diakonov:1997mm}. A weight diagram for anti-decuplet formed by using two flavor conjugate diquarks $(\bar{q})_{qq}$ and one anti-quark $\bar{q}$~\cite{Hosaka:2003jv}. Here $q\bar{q}$ denotes $u\bar{u}$, $d\bar{d}$, and $s\bar{s}$. The three corners of the triangle (in red) are exotic, which means that their quantum numbers require more than three quarks. 
} 
\label{fig:anti10} } }
\end{figure}

In the light-quark sector, Mitya, Vitya, and Maxim proposed a clearly exotic and narrow $\Theta^+ (uudd\bar{s})$ state~\cite{Diakonov:1986yh}, yet to be unequivocally observed and identified. The original name of the pentaquark lying at the apex of $\overline{\textbf{10}}$ was $Z^+$, then following Mitya's suggestion, we now call it $\Theta^+$ (see the Appendix). Using the Chiral Quark Soliton model~\cite{Diakonov:1986yh}, this particle along with other members of $\overline\textbf{{10}}$ has been proposed with a mass of $M_{\Theta^+} = 1.53~\mathrm{GeV}$ and a width less than $15~\mathrm{MeV}$~\cite{Diakonov:1997mm} (Fig.~\ref{fig:anti10}). Due to the relatively low mass and simple decay channels to $K^+n$ or $K^0p$, it has attracted the attention of many experiments at different facilities around the world (this renowned paper~\cite{Diakonov:1997mm}
has about 1000 citations collected by the Citation Index (CI) at iNSPIRE HEP).

The first experimental evidence for $\Theta^+$ came from the LEPS Collaboration at SPring-8 (investigating reaction $\gamma n\to K^+K^-n$ on $^{12}$C)~\cite{LEPS:2003wug} and the DIANA Collaboration from ITEP (investigating reaction $K^+n\to K^0p$ using data from the Xe bubble chamber)~\cite{DIANA:2003uet}.

The standard PWA (by construction) tends to miss a narrow resonance with $\Gamma_R < 
20~\mathrm{MeV}$ or so~\cite{Arndt:2003xz}. The modified PWA~\cite{Arndt:2003xz} assumes the existence of a narrow resonance by inserting a Breit-Wigner (BW) structure into one of the partial amplitudes and refitting the whole database. Such an analysis of the effect of the scanning mass $M_R$, full $\Gamma_R$ and partial width of the narrow resonance for each partial wave allows us to get a picture.  The worse description of the full database that includes resonance with the corresponding $M_R$ and $\Gamma_R$ (in terms of $\Delta\chi^2$) does not support our hypothesis. The better description may result in the resonance that may exist, or the effect can be due to various corrections (\textit{e.g.}, thresholds), or both possibilities can contribute. Some additional checks are necessary: true resonance should provide the effect only in a particular partial wave, while a non-resonance source may show similar effects in various partial waves.

\begin{figure}[htb!]
\centering
{
    \includegraphics[width=0.7\textwidth,keepaspectratio]{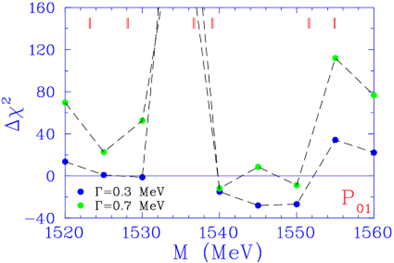} 
}

\centerline{\parbox{1\textwidth}{
\caption[] {\protect\small
Change of overall $\chi^2$ due to insertion of a resonance into $P_{01}$ for $M_R = 1520–1560~\mathrm{MeV}$ with $\Gamma_R = 0.3$ and $0.7~\mathrm{MeV}$, using KN plane wave 
approximation~\cite{Hyslop:1992cs}. Energies, where measurements exist, are labeled by red vertical bars (for references, see~\cite{Hyslop:1992cs}). The $1535~\mathrm{MeV}$ values for $\Delta\chi^2$ are off-scale~\cite{Arndt:2003xz}. Lines are given to guide the eye.} 
\label{fig:pwa1} } }
\end{figure}

Using this kind of modified PWA, Arndt and co-workers performed a reanalysis of the existing $KN$ database~\cite{Hyslop:1992cs} to refit $K^+N$ observables and consider the effect of a narrow
state~\cite{Arndt:2003xz}. The existence of a $\Theta^+$ in the $P_{01}$ ($J^P = 1/2^+$) state, with a mass of $\sim1545~\mathrm{MeV}$ and a width of $\Gamma(\Theta^+) \leq 0.5~\mathrm{MeV}$ or so, was found to be possible (Fig.~\ref{fig:pwa1}).

\section{My Story of $N(1680)$}
The mass difference in multiplet is determined by the phenomenological Gell-Mann-Okubo mass formula (mixing is able to shift some masses). Originally, Mitya, Vitya and Maxim claimed $N(1710)$ as a non-strange member of $\overline{\textbf{10}}$, which was a suitable candidate for $N^\ast$~\cite{Polyakov:2004}. The state $N(1710)$, though listed in the PDG2002 baryon summary
table~\cite{ParticleDataGroup:2002ivw} as a three-star resonance, was not seen in the PWA of the elastic scattering data $\pi N$ by the SAID group~\cite{Arndt:2003if}.

Studies that have claimed to see this $N(1710)$ state have given widely varying estimates of its mass (from $1680~\mathrm{MeV}$ to $1740~\mathrm{MeV}$) and width (from $90~\mathrm{MeV}$ to $500~\mathrm{MeV}$).  Branching ratios have also been given with large uncertainties (10–20\% for $N\pi$, 40–90\% for $N\pi\pi$, \textit{etc.}), apart from one that has been presented with greater precision ($6\pm 1\%$ for
$N\eta$)~\cite{ParticleDataGroup:2002ivw}.
\begin{figure}[htb!]
\centering
{
    \includegraphics[width=0.7\textwidth,keepaspectratio]{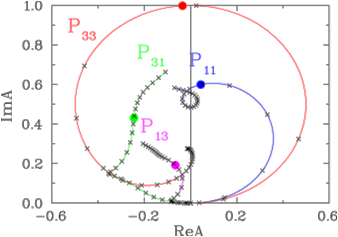} 
}

\centerline{\parbox{1\textwidth}{
\caption[] {\protect\small
Argand plots for the $P$ partial-wave amplitudes for the $\pi N$ case from threshold ($W = 1080~\mathrm{MeV}$) to $W = 2500~\mathrm{MeV}$~\cite{Arndt:2006bf}. The crosses indicate $50~\mathrm{MeV}$ steps in $W$. The filled circles correspond to the BW $W_R$.
} 
\label{fig:arg} } }
\end{figure}

The $\pi$N PWA by the GW SAID group has shown that above the Roper resonance~\cite{Roper:1964zza}, $W > 
1500~\mathrm{MeV}$, $\sigma^{tot} \cong 2\sigma^{el} \cong 2\sigma^{inel}$ ($\sigma^{tot} 
= \sigma^{el} + \sigma^{inel}$). It means that it is almost pure diffraction, inelasticity  
$\eta \to \infty$ for 
$P_{11}$ and $S \cong 0$ with $A \cong i/2$. The result is that phase $\delta$ is poorly defined and the $P_{11}$ amplitude is spinning around $ReA = 
0$~\cite{Arndt:2006bf} (Fig.~\ref{fig:arg}).

Of course, the nonobservation of a broad $N(1710)$ state in $\pi N$ elastic analyses (no pole position, no BW, no speed plot) could be due to a very small $\pi N$ branching ratio. The standard procedure used in PWA may also miss narrow resonances with $\Gamma < 20~\mathrm{MeV}$ by construction~\cite{Arndt:2003if}. Therefore, the true non-strange unitary partner of $\Theta^+$ (if it is different from $N(1710)$ and sufficiently narrow) could have eluded detection.

\begin{figure}[htb!]
\centering
{
    \includegraphics[width=0.7\textwidth,keepaspectratio]{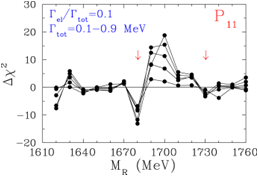} 
}

\centerline{\parbox{1\textwidth}{
\caption[] {\protect\small
Change of overall $\chi^2$ due to insertion of a resonance into $P_{11}$ for 
$M_R = 1660-1760~\mathrm{MeV}$ with $\Gamma_{tot} = 0.1\ (0.2)\ 0.9~\mathrm{MeV}$
and $\Gamma_{el} / \Gamma_{tot} = 0.1$ using $\pi N$ PWA~\cite{Arndt:2003ga}. Lines are
given to guide the eye. The red vertical arrows indicate $M_R = 1680~\mathrm{MeV}$ and $1730~\mathrm{MeV}$.} 
\label{fig:pwa2} } }
\end{figure}

Arndt and co-workers were considering the $\pi N$ partial wave $P_{11}$, as this is the amplitude that is associated with resonances having $J^P = 1/2^+$~\cite{Arndt:2003ga}. The character of
$\chi^2$ changes, $\Delta\chi^2$, after inserting a narrow resonance into the partial amplitude with a range of masses, widths, and branching fractions is illustrated in Fig.~\ref{fig:pwa2}. Negative values of $\Delta\chi^2$ emerge more readily near $M_R = 1680~\mathrm{MeV}$ and $1730~\mathrm{MeV}$. We see that $\Delta\chi^2$ becomes negative only for $\Gamma_{el} = (\Gamma_{el} / \Gamma_{tot})\cdot \Gamma_{tot}$ within the bounds $\Gamma_{el} \le 0.5~(0.3)~\mathrm{MeV}$ for $M_R = 1680~(1730)~\mathrm{MeV}$ (Fig.~\ref{fig:pwa2}). The available data cannot reliably discriminate values of $\Gamma_{el}$ below these bounds.

The expected decay properties of $N(1680)$ are essentially model-dependent and the Chiral Quark Soliton approach with violated SU(3)$_F$  [mixing $N_{\overline\textbf{{10}}} - N_\textbf{8}$] gives for $M_{N^\ast} = 1680~(1730)~\mathrm{MeV}$~\cite{Arndt:2003ga}:
\begin{itemize}
\item 
$\Gamma(N^\ast \to \pi\Delta) \approx 2.8~(3.5)~\mathrm{MeV}$.
\item 
$\Gamma(N^\ast \to K\Lambda) \approx 0.70~(1.56)~\mathrm{MeV}$.
\item 
$\Gamma(N^\ast \to \eta N) \approx 2~\mathrm{MeV}$.
\item 
$\Gamma(N^\ast \to \pi N) \approx 2.1~(2.3)~\mathrm{MeV}$.
\item 
$\Gamma(N^\ast \to \mathrm{tot}) \approx 10~\mathrm{MeV}$.
\end{itemize}

\begin{figure}[htb!]
\vspace{0.3cm}
\centering
{
    \includegraphics[width=0.8\textwidth,keepaspectratio]{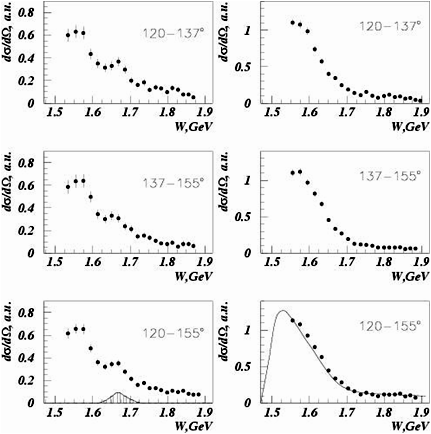} 
}

\centerline{\parbox{1\textwidth}{
\caption[] {\protect\small
Quasi-free $\eta n$ (left) and free $\eta p$ (right) photoproduction cross sections
(dark circles)~\cite{GRAAL:2004ndn}. The dashed area (left plot) shows the simulated contribution of a narrow state at $W = 1.675~\mathrm{MeV}$.
The solid line (right plot) indicates the solution of E429 of the SAID $\gamma p\to \eta p$ PWA~\cite{CrystalBallatMAMI:2010slt}. 
} 
\label{fig:graal} } }
\end{figure}

Then, the GRAAL Collaboration claimed evidence for a narrow ($10~\mathrm{MeV}$) resonance state with a mass of
$1675~\mathrm{MeV}$ in the reaction $\gamma n\to \eta n$ \\ (Fig.~\ref{fig:graal}~(left))~\cite{GRAAL:2004ndn}. BTW, in the case of a ``neutron'' target, the final-state interaction (FSI) is a problem. The effect is absent in the case of $\gamma p\to\eta p$ (Fig.~\ref{fig:graal}~(right)) and agrees with the prediction of the Chiral Quark Solution approach~\cite{Polyakov:2003dx}.  In exact SU(3)$_F$, the transition magnetic moment $\mu(p^\ast \to p)$ should vanish, since the U-spins are 3/2 for $p^\ast$, 1/2 for $p$, and 0 for the photon. With violation of SU(3)$_F$, this transition moment does not vanish but is still much smaller than $\mu(n^\ast \to n)$~\cite{Azimov:2006he}.

\begin{figure}[htb!]
\centering
{
    \includegraphics[width=0.522\textwidth,keepaspectratio]{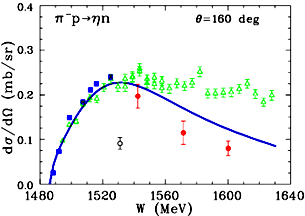}~~~
    \includegraphics[width=0.502\textwidth,keepaspectratio]{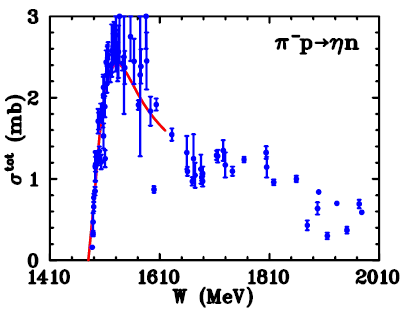} 
}

\centerline{\parbox{1\textwidth}{
\caption[] {\protect\small
\underline{Left}: Excitation function for differential cross sections for the reaction $\pi^-p\to \eta n$. Data came from
BNL, RHEL, and Saclay. Solid blue line from SAID PWA (WI08 solution)~\cite{Workman:2012hx}.
\underline{Right}: Total cross sections for the reaction $\pi^-p\to \eta n$. Solid red line from SAID PWA~\cite{Workman:2012hx}.
All data are available at Ref.~\cite{Briscoe:2020zzz}.} 
\label{fig:eta} } }
\end{figure}

The GW SAID group performed a coupled channel analysis of the $\pi N$ system that included elastic scattering of $\pi^\pm p$ and $\pi^-p\to \pi^0n$ with $\pi^-p \to \eta n$~\cite{Arndt:2003if}. The attractive factor for the reaction $\pi^-p \to \eta n$ is that it plays a role in isospin filtering. Unfortunately, the $\pi^-p \to \eta n$ data above $800~\mathrm{MeV}$ ($W = 1630~\mathrm{MeV}$) are not reliable for PWA (see, for instance, Fig.~\ref{fig:eta}~(left)). Most of the NIMROD data do not satisfy requirements (systematics is 10\% or more, momentum uncertainties are up to $100~\mathrm{MeV/c}$, and so on). That is the main reason why the SAID coupled channel analysis had no opportunity to look for $N(1680)$ using reaction data $\pi^-p \to \eta n$ (Fig.~\ref{fig:eta}~(right)).

%
%

EPECUR Collaboration at ITEP performed high quality measurements of the differential cross sections for elastic $\pi^\pm p$ over an energy range of $p = 800 - 1300~\mathrm{MeV/c}$ ($W = 1.55 - 1.83~\mathrm{GeV}$) and for center-of-mass angles from 40 to $120~\mathrm{deg}$~\cite{EPECUR:2014wrs}. In total, approximately 10,000 new data points have been obtained. These data have been produced with a momentum resolution of $\sim1~\mathrm{MeV/c}$ and with $\sim1$\% statistical uncertainties.
\begin{figure}[htb!]
\centering
{
    \includegraphics[width=0.6\textwidth,keepaspectratio]{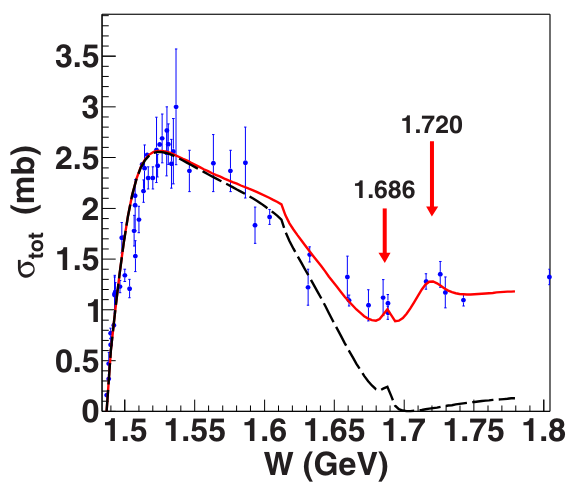} 
}

\centerline{\parbox{1\textwidth}{
\caption[] {\protect\small
Comparison of the measured $\pi$N data~\cite{Briscoe:2020zzz} and calculated $\pi^-p \to \eta n$ total cross sections. The solid red curve presents the calculations from the work~\cite{EPECUR:2016fmf}. The black dashed curve indicates the contribution of the $S$ wave.
}
\label{fig:gri} } }
\end{figure}

Following the analysis~\cite{Arndt:2003ga}, $\Gamma(\pi N) \sim 0.5~\mathrm{MeV}$, which stimulated us to look for $N(1680)$ using high-quality EPECUR elastic $\pi^\pm p$ scattering data~\cite{EPECUR:2016fmf}. The bottom line is that two narrow structures observed in elastic $\pi^-p$ scattering can be explained by a combination of threshold effects and two narrow resonances, $S_{11}(1686)$ and $P_{11}(1720)$. Fig.~\ref{fig:gri} presents the predictions for the $\pi^-p \to \eta n$ total cross sections. This model-dependent analysis shows that $S_{11}(1686)$ cannot be a member of $\overline{\textbf{10}}$, while $P_{11}(1720)$ may have a link to the results of Arndt \textit{et al.}~\cite{Arndt:2003ga}.

The new experimental proposal P102 for J-PARC addresses a measurement of the cross section of reaction $\pi^-p\to\eta n$ at $p = 850 - 1200~\mathrm{MeV/c}$ with a CsI (Tl) calorimeter~\cite{Kohri:2024}. The new data will help us solve the puzzle of $N(1680)$.

\section{Conclusion}
What would Mitya, Vitya, and Maxim think of the current situation in particle physics? On the one hand, they would be justifiably proud of the robust successes of an instanton-vacuum approach, which did so much to place it on a firm footing. However, some of the problems that concern them remain unresolved, notably the exotic $\Theta^+$ and its partners.

In fact, twenty years ago, in 2003, two experimental groups, LEPS and DIANA, announced the observation
of a light and narrow exotic baryon with mass in the vicinity of $1540~\mathrm{MeV}$, which was predicted by our friends. The history of this discovery and its theoretical interpretations can be found in the recent review~\cite{Praszalowicz:2024mji}.

Often memory does not keep details, and let me remind two events associated with
Mitya, Vitya, and Maxim and $\Theta^+$...

1) In 2009, I organized a Workshop on \textit{Narrow Nucleon Resonances: Predictions, Evidences, Perspectives}, Edinburgh, Scotland, June, 2009. \\
https://gwdac.phys.gwu.edu/~igor/Edinburgh2009/index.htm .
Unfortunately all three friends (Maxim was one of organizers) missed this event for some reasons, but many people attended and discussed physics.

2) Back in 2013, Nara, Japan, hosted the XVth International Conference on Hadron Spectroscopy (Hadron2013), Nara, Japan, in November 2013, which had a Memorial section for Mitya, https://pos.sissa.it/205/. Vitya was there. 
The contribution of Yakov Azimov and myself is in Ref.~\cite{Azimov:2013oxx}.

\begin{figure}[htb!]
\centering
{
    \includegraphics[width=0.6\textwidth,keepaspectratio]{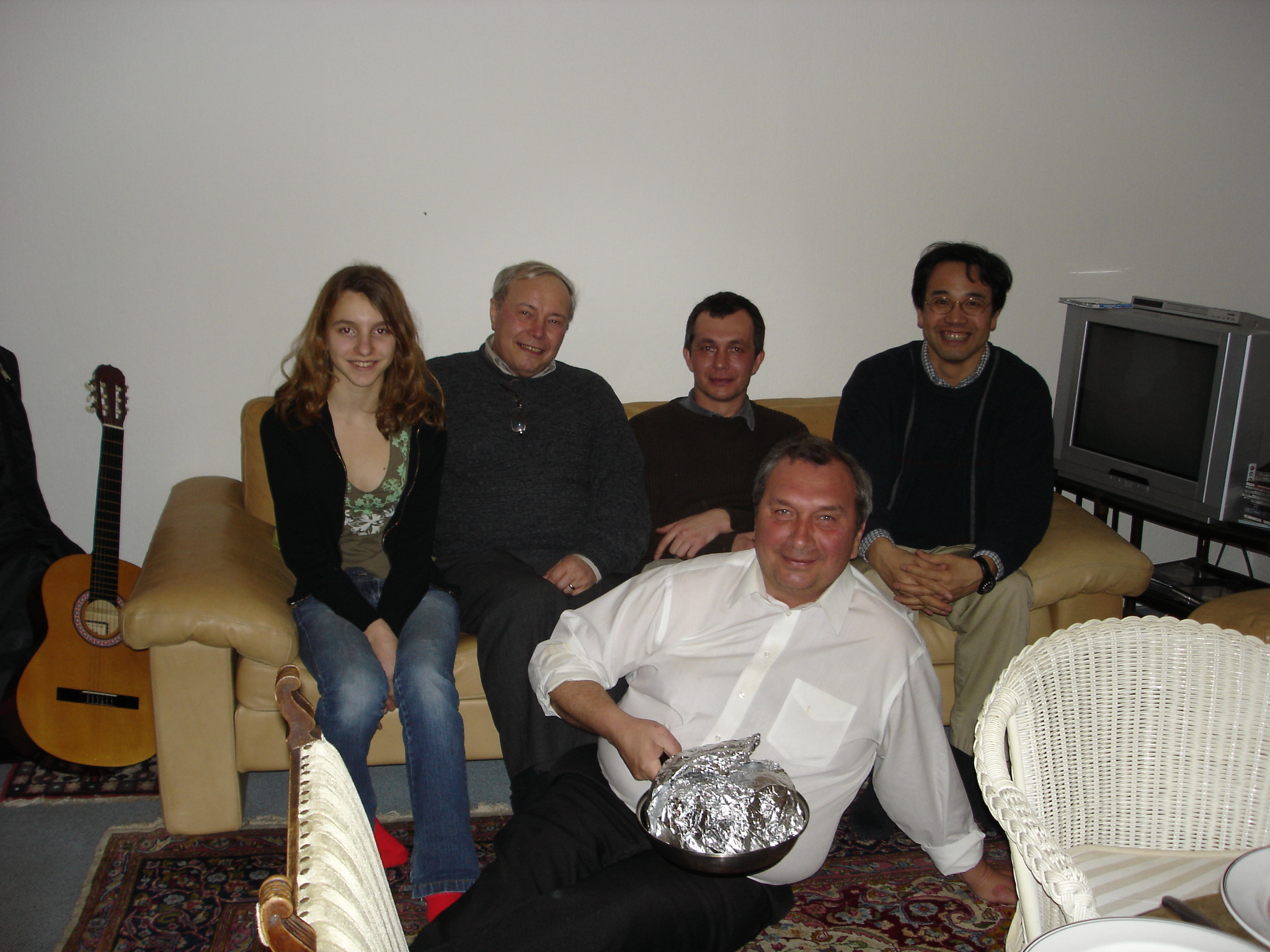} 
}

\centerline{\parbox{1\textwidth}{
\caption[] {\protect\small
Sitting on sofa from left to right: Masha, Maxim's daughter, me, Maxim, Takashi Nakano. Slava Kuznetsov is sitting just in front of us, he was a chef of our dinner.
}
\label{fig:photo} } }
\end{figure}

Finally, a photo taken in Maxim's apartment in Bochum, Germany, back to 2006 (Fig.~\ref{fig:photo}).

\section*{Acknowledgment}

This work was supported in part by the U.~S.~Department of Energy, Office of Science, Office of Nuclear Physics, under Award No.~DE--SC0016583.

\section*{Appendix}
My archive kept a letter for the name $\Theta^+$' that Mitya sent to the community~\cite{Diakonov:2003}. The result is that the community accepted Mitya's suggestion.\\

\textbf{From}:	Diakonov@nordita.dk \\
\textbf{To}: Takashi~Nakano, \\
    Anatoly~Dolgolenko,\\
    Elton~Smith,\\
    Valeri~Koubarovski,\\
    Ken~Hicks,\\
    Eugene~Pasyuk,\\
    Stepan~Stepanyan,\\
    Mark~Strikman,\\
    Maxim~Polyakov,\\
    Victor~Petrov,\\
    Igor~Strakovsky,\\
    James~Bjorken,\\
    Stan~Brodsky,\\
    Lev~B.~Okun,\\
    Leonid~G.~Landsberg,\\
    Hiroshi~Toki.\\
\textbf{Date}:	Apr 12, 2003, 12:33PM \\
\textbf{Subject}:	exotic baryon name \\

Dear friends,\\

Now that, thanks to your efforts, the exotic S=+1 baryon seems to become gradually a reality and not just a fancy dream of a theorist, we have to think how to name the child
to be born, hopefully. \\

The tentative name $Z^+$ we all have used is, to my mind, unfortunate. First, I have noticed that the first reaction by people hearing about it is that the new particle has some association with the electroweak Z boson. This association is completely misleading and may create lots of confusion in future and cause us much trouble. Second, the tradition is that baryons are named by capital Greek letters: Delta, Lambda, Sigma, Xi, Omega. (Except the nucleon but one cannot change it.) Third, ``Z'' is pronounced very differently in different languages; even in English it is ``zee'' in American English and ``zet'' in British which is mostly used in Europe and elsewhere. Fourth, and probably
most important, the name ``Z'' was imposed on us by the unsuccessful searches of exotic baryons in the far higher mass region, as summarized in the 1986 edition of Particle Data.
This was ``prehistoric'' time, before the discovery of the Z boson, and in any case nothing has been found worth mentioning in Particle Data Listings since the 1986 edition.
The $1540~\mathrm{MeV}$ signal has nothing to do with the broad and always suspicious one-star candidates in the $1700~\mathrm{MeV}$ region. I regret that we uncritically and unthoughtfully borrowed this queer name and used it in our paper with Petrov and Polyakov. But at that time it was so far from experiment! \\

I suggest that the name for this baryon should satisfy the following criteria: \\

1) It must be a capital Greek letter, according to the tradition of naming baryons. \\

2) It must be distinct from anything used before and carry no associations with bosons. \\

3) Last but not least, the character must exist in LaTeX. \\

If you look into the list of upper case Greek letters used in LaTeX you'll find that there is one and only one character satisfying all criteria, and it is \\

THETA. \\

When I discovered it I realized that I kind of liked this name. It is a symmetric and ``round'' character like the Omega also sitting at the vertex of the big decuplet triangle and, like Omega, alluding to that it is a singlet. If you disregard the historic ``tau-theta problem'' of the last century's 50s, this character has been never used in particle physics, as far as I know. It carries no associations and hints that it is something really new, which it certainly is. Last, ``Theta'' is pronounced more or less in the same way in all languages I know. \\

Therefore, I suggest that we, experimentalists and theorists, will henceforth call the new candidate for the exotic baryon with strangeness +1 and mass around $1540~\mathrm{MeV}$. \\

$\Theta^+$ baryon. \\

Any objections? \\

With my best regards, \\
Dmitri Diakonov.



\begin{thebibliography}{99}
\bibitem{Diakonov:1997mm}
   D.~Diakonov, V.~Petrov, and M.~V.~Polyakov,
   ``Exotic anti-decuplet of baryons: Prediction from chiral solitons,''
   Z.\ Phys.\ A\ \textbf{359}, 305 (1997).
\bibitem{Hyslop:1992cs}
   J.~S.~Hyslop, R.~A.~Arndt, L.~D.~Roper, and R.~L.~Workman,
   ``Partial wave analysis of $K^+$-nucleon scattering,''
   Phys.\ Rev.\ D\ \textbf{46}, 961 (1992).
\bibitem{Arndt:2003xz}
   R.~A.~Arndt, I.~I.~Strakovsky, and R.~L.~Workman,
   ``$K^+$ nucleon scattering and exotic S = +1 baryons,''
   Phys.\ Rev.\ C\ \textbf{68}, 042201 (2003);
   [erratum: Phys.\ Rev.\ C\ \textbf{69}, 019901 (2004)].
\bibitem{Pavan:2001wz}
    M.~M.~Pavan, I.~I.~Strakovsky, R.~L.~Workman, and R.~A.~Arndt,
    ``The pion nucleon sigma term is definitely large: Results from a GWU analysis of $\pi$-nucleon scattering data,''
    $\pi$N\ Newslett.\ \textbf{16}, 110 (2002);
    [arXiv:hep-ph/0111066 [hep-ph]].
\bibitem{Diakonov:1985eg}
    D.~Diakonov and V.~Y.~Petrov,
    ``A theory of light quarks in the Instanton Vacuum,''
    Nucl.\ Phys.\ B\ \textbf{272}, 457 (1986).
\bibitem{Diakonov:1988ju}
    D.~Diakonov, V.~Y.~Petrov, and P.~V.~Pobylitsa,
    ``Born diagrams in the pion - Skyrmion scattering,''
    Phys.\ Lett.\ B\ \textbf{205}, 372 (1988).
\bibitem{Gell-Mann:1964ewy}
   M.~Gell-Mann,
   ``A schematic model of baryons and mesons,''
   Phys.\ Lett.\ \textbf{8}, 214 (1964).
\bibitem{ParticleDataGroup:2024cfk}
    S.~Navas \textit{et al.} [Particle Data Group],
    ``Review of Particle Physics,''
    Phys.\ Rev.\ D\ \textbf{110}, 030001 (2024).
\bibitem{LHCb:2019kea}
   R.~Aaij \textit{et al.} [LHCb Collaboration],
   ``Observation of a narrow pentaquark state, $P_c(4312)^+$, and of two-peak structure of the $P_c(4450)^+$,''
   Phys.\ Rev.\ Lett.\ \textbf{122}, 222001 (2019).
\bibitem{Eides:2017xnt}
    M.~I.~Eides, V.~Y.~Petrov, and M.~V.~Polyakov,
    ``Pentaquarks with hidden charm as hadroquarkonia,''
    Eur.\ Phys.\ J.\ C\ \textbf{78}, 36 (2018).
\bibitem{Diakonov:1986yh}
    D.~Diakonov and V.~Y.~Petrov,
    ``Chiral theory of nucleons,''
    Pisma\ Zh.\ Eksp.\ Teor.\ Fiz.\ \textbf{43}, 57 (1986)
    [JETP\ Lett.\ \textbf{43}, 75 (1986)].
\bibitem{Hosaka:2003jv}
    A.~Hosaka,
    ``Pentaquark states in a chiral potential,''
    Phys.\ Lett.\ B\ \textbf{571}, 55 (2003).
\bibitem{LEPS:2003wug}
   T.~Nakano \textit{et al.} [LEPS Collaboration],
   ``Evidence for a narrow S = +1 baryon resonance in photoproduction from the neutron,''
   Phys.\ Rev.\ Lett.\ \textbf{91}, 012002 (2003).
\bibitem{DIANA:2003uet}
   V.~V.~Barmin \textit{et al.} [DIANA Collaboration],
   ``Observation of a baryon resonance with positive strangeness in collisions $K^+$ with Xe nuclei,''
    Yad.\ Fiz.\ \textbf{ 66}, 1763 (2003) 
    [Phys.\ Atom.\ Nucl.\ \textbf{66}, 1715 (2003)].
\bibitem{Polyakov:2004}
   M.~V.~Polyakov (private communication).
\bibitem{ParticleDataGroup:2002ivw}
   K.~Hagiwara \textit{et al.} [Particle Data Group],
   ``Review of Particle Physics,''
   Phys.\ Rev.\ D\ \textbf{66}, 010001 (2002).
\bibitem{Arndt:2003if}
   R.~A.~Arndt, W.~J.~Briscoe, I.~I.~Strakovsky, R.~L.~Workman, and M.~M.~Pavan,
   ``Dispersion relation constrained partial wave analysis of $\pi N$ elastic and $\pi N\to \eta N$ scattering data: The Baryon spectrum,''
   Phys.\ Rev.\ C\ \textbf{69}, 035213 (2004).
\bibitem{Roper:1964zza}
    L.~D.~Roper,
    ``Evidence for a P-11 pion-nucleon resonance at 556~MeV,''
    Phys.\ Rev.\ Lett.\ \textbf{12}, 340 (1964).
\bibitem{Arndt:2006bf}
    R.~A.~Arndt, W.~J.~Briscoe, I.~I.~Strakovsky, and R.~L.~Workman,
    ``Extended partial-wave analysis of $\pi N$ scattering data,''
    Phys.\ Rev.\ C\ \textbf{74}, 045205 (2006).
\bibitem{Arndt:2003ga}
   R.~A.~Arndt, Y.~I.~Azimov, M.~V.~Polyakov, I.~I.~Strakovsky, and R.~L.~Workman,
   ``Nonstrange and other unitarity partners of the exotic $\Theta^+$ baryon,''
   Phys.\ Rev.\ C\ \textbf{69}, 035208 (2004).
\bibitem{GRAAL:2004ndn}
   V.~Kuznetsov \textit{et al.} [GRAAL Collaboration],
   ``$\eta$ photoproduction off the neutron at GRAAL: Evidence for a resonant structure at W = 1.67-GeV,''
   [arXiv:hep-ex/0409032 [hep-ex]].
\bibitem{Polyakov:2003dx}
   M.~V.~Polyakov and A.~Rathke,
   ``On photoexcitation of baryon anti-decuplet,''
   Eur.\ Phys.\ J.\ A\ \textbf{18}, 691 (2003).
\bibitem{Azimov:2006he}
   Y.~I.~Azimov, V.~Kuznetsov, M.~V.~Polyakov, and I.~Strakovsky,
   ``K*-couplings for the antidecuplet excitation,''
   Phys.\ Rev.\ D\ \textbf{75}, 054014 (2007).
\bibitem{CrystalBallatMAMI:2010slt}
   E.~F.~McNicoll \textit{et al.} [Crystal Ball at MAMI Collaboration],
   ``Study of the $\gamma p \to \eta p$ reaction with the Crystal Ball detector at the Mainz Microtron (MAMI-C),''
   Phys.\ Rev.\ C\ \textbf{82}, 035208 (2010)
   [erratum: Phys.\ Rev.\ C\ \textbf{84}, 029901 (2011)].
\bibitem{Briscoe:2020zzz}
  The SAID analyses are available through the GWU web-site: http://gwdac.phys.gwu.edu/;  
  W.~J.~Briscoe, M.~D\"oring, H.~Haberzettl, I.~I.~Strakovsky, and R.~L.~Workman,
  Institute of Nuclear Studies of The George Washington University Database.
\bibitem{Workman:2012hx}
   R.~L.~Workman, R.~A.~Arndt, W.~J.~Briscoe, M.~W.~Paris, and I.~I.~Strakovsky,
   ``Parameterization dependence of T matrix poles and eigenphases from a fit to $\pi$N elastic scattering data,''
   Phys.\ Rev.\ C\ \textbf{86}, 035202 (2012).
\bibitem{EPECUR:2014wrs}
   I.~G.~Alekseev \textit{et al.} [EPECUR Collaboration],
   ``High-precision measurements of $\pi p$ elastic differential cross sections in the second resonance region,''
  Phys.\ Rev.\ C\ \textbf{91}, 025205 (2015).
\bibitem{EPECUR:2016fmf}
   A.~Gridnev \textit{et al.} [EPECUR Collaboration],
   ``Search for narrow resonances in $\pi p$ elastic scattering from the EPECUR experiment,''
   Phys.\ Rev.\ C\ \textbf{93}, 062201 (2016).
\bibitem{Kohri:2024}
   H.~Kohri \textit{et al.} [P102 Collaboration], ``Study of the peculiar bump structure at $1680~\mathrm{MeV}$ by the $\pi^-p\to \eta n$ reaction with momenta of $p_\pi = 0.85-1.2~\mathrm{GeV/c}$,'' J-PARC Experimental Proposal P102 (2023);\\
   https://j-parc.jp/researcher/Hadron/en/pac$_-$2401/pdf/P102$_-$2024-4.pdf.
\bibitem{Praszalowicz:2024mji}
    M.~Praszalowicz,
    ``Odyssey of the elusive $\Theta^+$,''
    [arXiv:2411.08429 [hep-ph]].
\bibitem{Azimov:2013oxx}
    Y.~Azimov and I.~Strakovsky,
    ``Exotics and PWA for $\pi$N Scattering,''
    PoS \textbf{Hadron2013}, 034 (2013).
\bibitem{Diakonov:2003}
    D.~I.~Diakonov (private communication).
\end{thebibliography}
\end{document}